\documentclass[a4paper,fleqn,usenatbib, hyperref]{mnras}

\usepackage{comment}
\usepackage[T1]{fontenc}
\usepackage{ae,aecompl}
\usepackage{hyperref}
\usepackage{graphicx}
\usepackage{mathrsfs}
\usepackage{amsmath}
\usepackage{amssymb}
\usepackage{verbatim}
\usepackage{multirow}
\usepackage{subfig}
\usepackage{mathtools}
\usepackage[noabbrev]{cleveref}
\usepackage{eqnarray,amsmath}
\def\apjl{ApJ}
\def\apj{ApJ}                 
\def\apjl{ApJ}                
\def\aap{A\&A}                
\def\mnras{MNRAS}             
\def\nat{Nature}              

\def \inte {{\em INTEGRAL}}

\def \nicer {{\em NICER}}

\def \saxj{SAX J1808.4$-$3658}

\def \igr{IGR J17591$-$2342}

\title[\igr{}: spin-down during accretion]{Timing of the accreting millisecond pulsar \igr{}: evidence of spin-down during accretion}
\author[Sanna et al. ]{A. Sanna$^{1,3}$\thanks{E-mail:
    andrea.sanna@dsf.unica.it}, L. Burderi$^{1,2,3}$, K.~C. Gendreau$^{4}$, T. Di Salvo$^{2,3,5}$, P.~S. Ray$^{6}$,\newauthor  A. Riggio$^{1,3}$, A. F. Gambino$^{5}$, 
     R. Iaria$^{5}$, L. Piga$^{1}$, C.~Malacaria$^{7,8}$, G.~K. Jaisawal$^{9}$\\
$^{1}$Dipartimento di Fisica, Universit\`a degli Studi di Cagliari, SP Monserrato-Sestu km 0.7, 09042 Monserrato, Italy\\
$^{2}$INFN, Sezione di Cagliari, Cittadella Universitaria, 09042 Monserrato, CA, Italy\\
$^{3}$INAF - Osservatorio Astronomico di Cagliari, via della Scienza 5, 09047 Selargius (CA), Italy\\
$^{4}$Astrophysics Science Division, NASA's Goddard Space Flight Center, Greenbelt, MD 20771, USA\\
$^{5}$Universit\`a degli Studi di Palermo, Dipartimento di Fisica e Chimica, via Archirafi 36, 90123 Palermo, Italy\\
$^{6}$Space Science Division, Naval Research Laboratory, Washington, DC 20375-5352, USA\\
$^{7}$NASA Marshall Space Flight Center, NSSTC, 320 Sparkman Drive, Huntsville, AL 35805, USA\\
$^{8}$Universities Space Research Association, Science and Technology Institute, 320 Sparkman Drive, Huntsville, AL 35805, USA\\
$^{9}$National Space Institute, Technical University of Denmark, Elektrovej 327-328, 2800 Lyngby, Denmark
}
\begin{document}

\date{Accepted -. Received -; in original form -}

\pagerange{\pageref{firstpage}$-$\pageref{lastpage}} \pubyear{2020}

\maketitle

\label{firstpage}

\begin{abstract}

We report on the phase-coherent timing analysis of the accreting millisecond X-ray pulsar \igr{}, using Neutron Star Interior Composition Explorer (\nicer{}) data taken during the outburst of the source between 2018 August 15 and 2018 October 17. We obtain an updated orbital solution of the binary system. We investigate the evolution of the neutron star spin frequency during the outburst, reporting a refined estimate of the spin frequency and the first estimate of the spin frequency derivative ($\dot{\nu}\sim -7\times10^{-14}$ Hz s$^{-1}$), confirmed independently from the modelling of the fundamental frequency and its first harmonic. We further investigate the evolution of the X-ray pulse phases adopting a physical model that accounts for the accretion material torque as well as the magnetic threading of the accretion disc in regions where the Keplerian velocity is slower than the magnetosphere velocity. From this analysis we estimate the neutron star magnetic field $B_{eq}=2.8(3)\times 10^8$ G. Finally, we investigate the pulse profile dependence on energy finding that the observed behaviour of the pulse fractional amplitude and lags as a function of energy are compatible with a thermal Comptonisation of the soft photons emitted from the neutron star caps.
\end{abstract}


\section{Introduction}

One of the pillars of accretion theory onto Neutron Stars (NS, hereinafter) by a low mass companion ($\lesssim 1 {\rm M_{\odot}}$) overflowing its Roche Lobe is the recognition that the accreting matter dumps its specific angular momentum onto the NS, causing the spin of the compact object ($\nu$) to vary in response to the accretion process. This prediction is based onto the very robust assumption that angular momentum of an isolated system (the NS in this case) is a conserved quantity and therefore the evolution of its angular momentum is determined by the balance of all the torques acting on it. Based on that, accreting millisecond X-ray pulsars (AMXPs) have been interpreted as the result of long-lasting mass accretion onto an old slow-rotating pulsar from an evolved companion star \cite[see][for more details on the recycling scenario]{Alpar82}. The evolutionary link between accreting low-mass X-ray binaries and the rotation powered millisecond radio pulsars (MSP) has been observationally proven in a few systems called transitional MSP \citep[see e.g.][]{Archibald2015a, Papitto2013b,Bassa2014a,Papitto2015a} where a transition from the radio MSP phase (rotation powered) to the X-ray AMXP phase (accretion powered) has been detected.

Twenty-two AMXPs are currently known \citep[for an extensive review, see e.g.][]{Patruno12b,Campana2018a}. Simple considerations on the torque exerted on NS in AMXPs by accreting matter would suggest an increase of their frequency (spin-up) during the outburst whilst they should spin-down during quiescence phases. Interesting, studies performed on AMXPs observed in outburst led to controversial results. Indeed, while a subset of systems observed during accretion (see e.g. IGR J00291+5934, XTE J1807-294,  XTE J1751-305, IGR J17511-3057), clearly showed an increase of the spin frequency with time \citep{Falanga05b, Riggio08,Papitto08,Riggio11b}, others sources (see e.g. XTE J0929-314, XTE J1814-338, IGR J17498-2921 and SAX J1808.4-3658) show spin-down during outbursts \citep{Galloway02,Burderi06,Papitto07, Papitto11c, Bult19}. Braking torques on NSs, generated by the interaction of the NS (dipolar) magnetic field threading the accretion disc outside the co-rotation radius, has been proposed to explain negative spin derivatives during the accretion phases of AMXPs \citep[see e.g.][and references therein]{Wang87, Rappaport04, Kluzniak2007}. On the other hand,
estimates of the NS spin derivatives are difficult to obtain as coherent timing analysis has been proven to be highly sensitive to X-ray timing noise, most probably related to variations of the X-ray flux \citep[see e.g. ][]{Patruno09c}, which can hinder the detection of relative weak spin derivatives \citep[see e.g.][]{Galloway02, Burderi06, Riggio08, Bult19}.

Here, we report on the phase-coherent timing analysis of the AMXP \igr{} \citep{Sanna:2018ab}, firstly detected by \inte{} on August 10, 2018 \citep{Ducci2018aa} and extensively monitored by the \nicer{} instrument during its outburst. Moreover, we investigated the evolution of the spin pulse delays within the framework of magnetically disc-threading theories in which the material torque (dependent on mass accretion rate) as well as the disc-threading torques exerted by the magnetic field are taken into account \citep[see e.g.][]{Rappaport04, Kluzniak07}. For the first time in phase-coherent timing analysis of AMXPs, we modelled variations of the frequency spin derivative taking into account instantaneous values of the mass accretion rate by following the X-ray flux evolution. Recently, a similar approach has been successfully applied to the timing of the X-ray pulsar GRO J1744-28, a slowly spinning ($\sim 2$ Hz) NS accreting from a low-mass companion \citep{Sanna:2017aa}. The obtained results show that, at least in this case, the standard theory of accretion onto NS works very well also for fast rotators and allowed an accurate determination of the magnetic field strength that is in the range of the values expected for this class of sources. Finally, we report on the updated source ephemerides and we discuss the properties of the X-ray pulsation as well as its temporal evolution.

\section[]{Observations and data analysis}
\label{sec:tim}
\subsection{\nicer{}}
\begin{figure}
\centering
\includegraphics[width=0.48\textwidth]{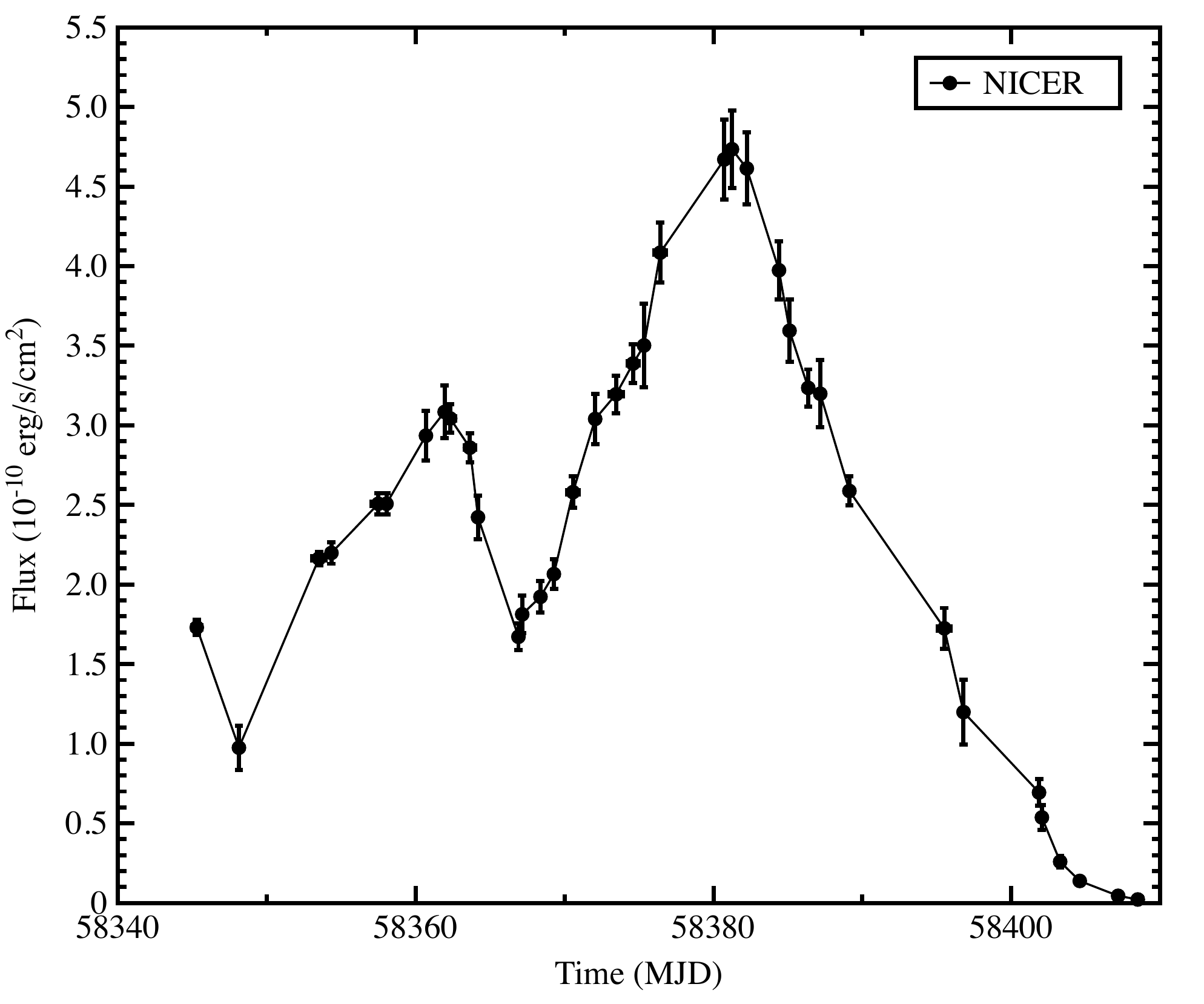}
\caption{0.5--10 keV unabsorbed flux of \igr{}, as taken by
the \nicer{} during the entire outburst. Each point represents a single observation of the source modelled with \textit{Tbabs(bbody+nthcomp)} in Xspec.}
\label{fig:flux}
\end{figure}

\nicer{} \citep{Gendreau2012} started monitoring the X-ray transient \igr{} on 2018 August 15 (MJD 58345) up to 2018 October 17 (MJD 58408.2) for a total of 37 dedicated observations. We processed the available \nicer{} observation using the \textsc{nicerdas} pipeline (version V004a) retaining events in the 0.2--12keV energy range, for which the pointing offset was $< 54''$, the dark Earth limb angle was $> 30^\circ$, the bright Earth limb angle was  $> 40^\circ$, and the ISS location was outside of the South Atlantic Anomaly (SAA). Moreover, we removed short intervals that showed background flaring in high geomagnetic latitude regions. As a result of the filtering process we retained a total exposure time of  $\sim101$ ks. The spectral background was obtained from a blank-sky observations of the RXTE-6 region. We did not detect any Type-I thermonuclear bursts during the observations analysed in this work. For each observation we modelled the energy spectrum with an absorbed black body component plus a thermally comptonised continuum (\textit{Tbabs(bbody+nthcomp)} in Xspec) and we extracted the source unabsorbed flux in the energy range 0.5--10 keV (see Fig.~\ref{fig:flux}). 
Finally, we barycentred the \nicer{} photon arrival times with the \textsc{BARYCORR} tool using the JPL DE-405 Solar System ephemeris and adopting the source coordinates obtained  from the ATCA radio detection of the source \citep{Russell2018aa}.

\begin{figure*}
\centering
\includegraphics[width=0.8\textwidth]{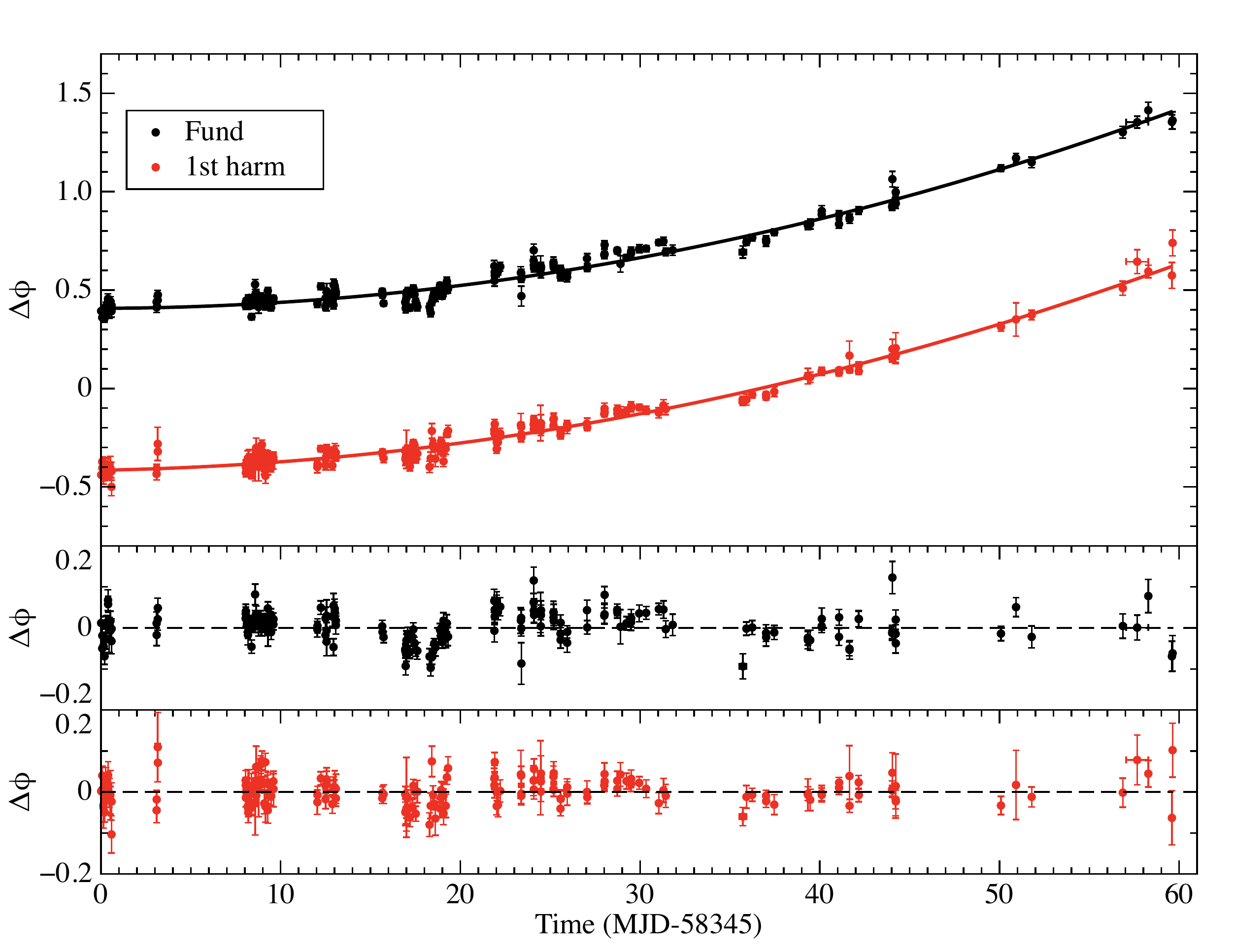}
\caption{\textit{Top panel -} Time evolution of the phase delays for the fundamental (black dots) and the first harmonic (red dots) components used to model the source pulse profiles created by epoch-folding segments of \nicer{} data with duration between 200 and 1500 seconds. The solid lines represent the best-fitting models for the two components.  \textit{Middle panel -} Fundamental residuals in pulse phase units with respect to its best-fitting solution. \textit{Bottom panel -} first harmonic residuals in pulse phase units with respect to its best-fitting solution.}
\label{fig:phase_fit}
\end{figure*}

\begin{figure}
\centering
\includegraphics[width=0.48\textwidth]{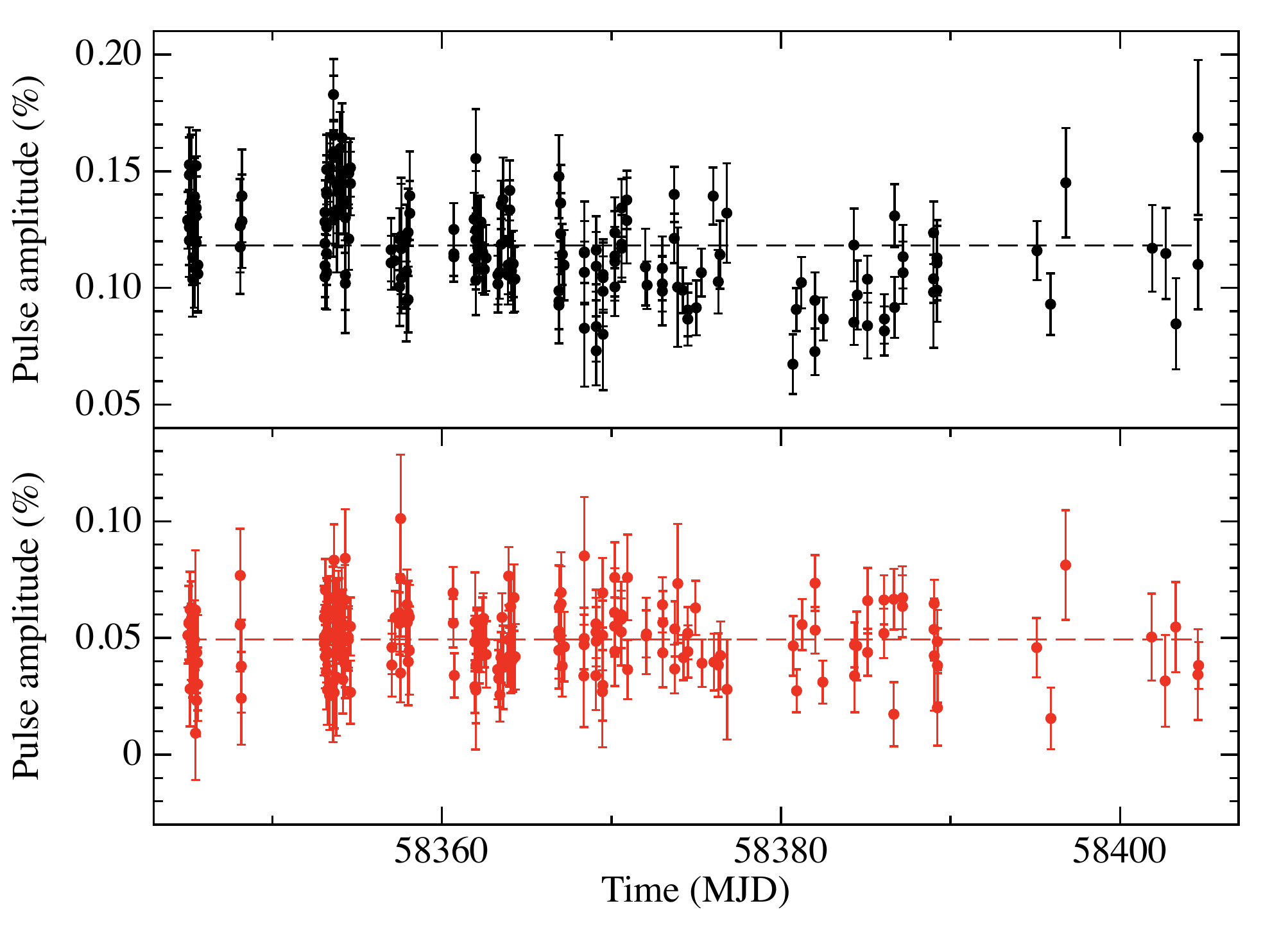}
\caption{Time evolution of the fractional amplitudes of the two harmonic components (top panel shows to the fundamental frequency, while bottom panel the first harmonic) adopted to model the source pulse profiles.}
\label{fig:amp}
\end{figure}

To perform phase-coherent timing analysis we started by correcting the \nicer{} photon times of arrival for the binary Doppler delay estimated assuming an almost circular orbits \citep[see e.g.][]{Deeter81,Burderi07,Sanna2016a} with parameters equal to that reported by \citet{Sanna:2018ab} obtained from the analysis of a small portion of the source outburst (almost 10 out 60 days).  We created pulse profiles epoch-folding the available data at the previously determined spin frequency $\nu_0=527.42570042$ Hz. To optimise the statistics and increase the number of profiles, we split the datasets in segments with length ranging between 200 and 1500 seconds. We modelled the pulse profile as the combination of two harmonically related sinusoidal functions: the fundamental and the first harmonic components of the spin frequency. To proceed further with the analysis, we selected only statistically significant pulse profiles, i.e. profiles for which the ratio between the sinusoidal amplitude and its 1$\sigma$ uncertainty was greater than three. As shown in Fig.~\ref{fig:phase_fit}, coherent oscillations were clearly detected in the time interval MJD 58345 -- MJD 58405, covering almost entirely the outburst duration. The fractional amplitude of the fundamental and the first harmonic components stays overall constant around a mean value of $\sim$11.8\% and $\sim$5\%, respectively (Fig.~\ref{fig:amp})

Following standard timing techniques \cite[see e.g.][]{Burderi07,Sanna:2018ab}, we modelled the temporal evolution of the pulse phase delays obtained from the fundamental and first harmonic components as follows:
\begin{equation}
\label{eq:ph}
\Delta \phi(t)=\phi_0+\Delta \nu_0\,(t-T_0)-\frac{1}{2}\dot{\nu}\,(t-T_0)^2+R_{orb}(t)
\end{equation}
where $\Delta \nu_0=(\nu_0-\bar{\nu})$ and $\dot{\nu}$ represent the spin frequency correction and the spin frequency derivative, respectively, estimated with respect to the reference epoch $T_0$. $R_{orb}$ models the differential corrections to the ephemeris used to generate the pulse phase delays \citep[see e.g.][]{Deeter81}. To improve the accuracy on the orbital parameters we modelled simultaneously pulse phase delays from the fundamental and first harmonic components using Eq.~\ref{eq:ph}. More specifically, we fitted the pulse phases linking the $R_{orb}(t)$ component, meaning that the orbital parameters will be the same during the fit of the two datasets. We applied iteratively the method, until no significant differential corrections were found for the parameters of the model. In the right column of Tab.~\ref{tab:solution}, we report the the best-fit parameters, while in Fig.~\ref{fig:phase_fit} we show the pulse phase delays with the best-fitting model (top panel; black and red points for the fundamental and first harmonic components, respectively), and the corresponding residuals (in phase units) with respect to the best-fitting model to fit the time evolution of the fundamental (middle panel) and first overtone (bottom panel) phase delays, respectively. The value of $\tilde{\chi}^2\sim2.5$ (with 355 degrees of freedom), as well as the distribution of the residuals suggest the presence of timing noise, largely observed in several AMXPs \cite[see e.g.][]{Burderi06,Papitto07,Riggio08,Riggio11a}. We rescaled the uncertainties on the parameters reported in Tab.~\ref{tab:solution} by a factor $\sqrt{\chi^2}$ to give a realistic estimation of the uncertainties given the relatively poor fit of the pulse phases \citep[see e.g.][]{Finger1999}.

\begin{table*}

\begin{tabular}{l | c  c  }
\hline
Parameters             & S18 & this work  \\
\hline
\hline
R.A. (J2000) &  \multicolumn{2}{c}{$17^h59^m02^s.86\pm0.04^s$}\\
Decl. (J2000) & \multicolumn{2}{c}{$-23^\circ43'08''.3\pm0.1''$}\\
Orbital period $P_{orb}$ (s) & 31684.743(3)&31684.7503(5)\\
Projected semi-major axis $a \sin i/c$ (lt-s) &1.227716(8)&1.227714(4) \\
Ascending node passage $T_{\text{NOD}}$ (MJD) & 58345.1719787(16) & 58345.1719781(9)\\
Eccentricity ($e$) &< 6$\times 10^{-5}$ & < 5$\times 10^{-5}$\\
$\chi^2$/d.o.f. &123.75/99 & 876.4/355\\
\hline
\multicolumn{3}{c}{Fundamental}\\
\hline
\hline
Spin frequency $\nu_0$ (Hz) &527.42570042(8)& 527.425700578(9)\\
Spin frequency 1st derivative $\dot{\nu}_0$ (Hz/s) &2.0(1.6)$\times 10^{-13}$&$-7.4$(4)$\times 10^{-14}$\\
\hline
\multicolumn{3}{c}{1st Harmonic}\\
\hline
\hline
Spin frequency $\nu_0$ (Hz) &--& 527.42570056(1)\\
Spin frequency 1st derivative $\dot{\nu}_0$ (Hz/s)&--&$-7.1$(4)$\times 10^{-14}$\\
\hline
\end{tabular}
\caption{Orbital ephemeris of \igr{} obtained from the timing analysis of the \nicer{} observations collected during the whole outburst of the source. The orbital solution is referred to the epoch T$_0$=58345.0 MJD. Errors are at 1$\sigma$ confidence level.}
\label{tab:solution}
\end{table*}

To explore the effect of the positional uncertainties on the spin frequency and spin frequency derivative reported in Tab.~\ref{tab:solution}, we considered the residuals induced by the motion of the Earth for small variations of the source position $\delta_{\lambda}$ and $\delta_{\gamma}$ (ecliptic coordinates) expressed by the relation:
\begin{equation}
R_{pos}(t) = - \nu_0 y [\sin(M_0+\epsilon)\cos \gamma \delta\lambda -  \cos(M_0+\epsilon)\sin \gamma \delta\gamma],
\label{eq:pos}
\end{equation}
where $y=r_E/c$ represents the Earth semi-major axis in light-seconds, $M_0=2 \pi (T_0-T_{v})/P_{\oplus}-\lambda$, $T_{v}$ and $P_{\oplus}$ are the vernal point and the Earth orbital period, respectively, and $\epsilon=2\pi(t-T_0)/P_{\oplus}$ \citep[see, e.g.][]{Lyne90}. Degeneracy between phase delay residuals induced by positional uncertainties and those related to spin frequency uncertainties and/or spin frequency derivative is unavoidable on timescales much shorter (i.e. 60 days outburst duration) than Earth's orbital period. Upper limits on these quantities can be obtained expanding Eq.\ref{eq:pos} in series of $\epsilon\ll1$ \cite[see e.g.][and references therein]{Burderi06}. More specifically, the systematic error on the spin frequency correction and the spin frequency derivative are $\sigma_{\nu_{pos}}\leq \nu_0y\sigma_{v}(1+\sin^2\gamma)^{1/2}2\pi/P_{\oplus}$ and $\sigma_{\dot{\nu}_{pos}}\leq \nu_0y\sigma_{v}(1+\sin^2\gamma)^{1/2}(2\pi/P_{\oplus})^2$, respectively, where $\sigma_{v}$ is the positional error circle. Considering the uncertainty on the source position reported in Tab.~\ref{tab:solution} \citep{Russell2018aa}, we estimated $\sigma_{\nu_{pos}} \leq 7\times 10^{-9}$ Hz and $\sigma_{\dot{\nu}_{pos}} \leq 2\times 10^{-15}$ Hz/s, respectively. These systematic uncertainties, compatible with the statistical uncertainties reported in this work, have been added in quadrature to the statistical errors of $\nu_0$ and $\dot{\nu}$ estimated from the timing analysis.

We carried out studies on the properties of the pulse profile as a function of energy by dividing the \nicer{} energy range between 0.2 keV to 12 keV into 20 intervals. Each background-subtracted pulse profile has been modelled as the superposition of two harmonically related sinusoidal functions (fundamental and first harmonic) for which we calculated the fractional amplitudes and the time lags with respect to the lowest energy selection. In the top panel of Fig.~\ref{fig:amp_vs_energy} we report the fractional amplitude as a function of energy for the fundamental (black points) and first harmonic (red points) pulse phase component of the pulse profile. The fractional amplitude of the fundamental component increases from $\sim4\%$ up to $\sim16\%$ at around 5 keV, decreasing then to $\sim12\%$ at 12 keV. A similar behaviour, but limited between  $\sim2.6\%$ and $\sim6.6\%$ is shown by the first harmonic. The bottom panel of Fig.~\ref{fig:amp_vs_energy} suggests the presence of hard lags (pulsed high energy photons are observed earlier with respect to soft ones) almost linearly increasing with energy for the two components. Interestingly, while the first harmonic component only shows soft lags in the energy range explored, the fundamental component shows no lags or marginally hard lags between 0.5--3.5 keV followed by soft lags above 3.5 keV.

\begin{figure}
\centering
\includegraphics[width=0.48\textwidth]{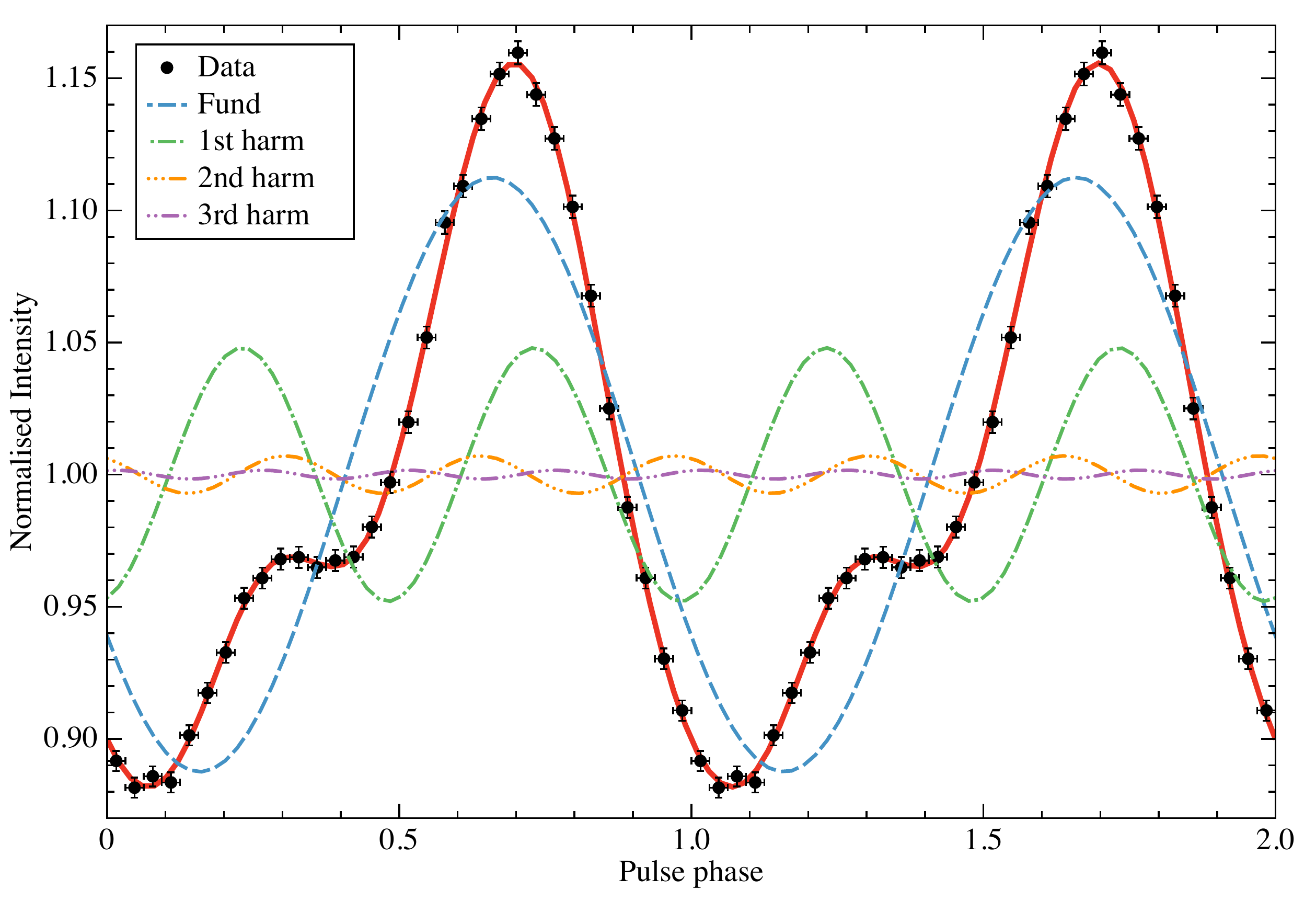}
\caption{\igr{} pulse profile (black points) obtained epoch-folding the whole available \nicer{} dataset after correcting for the orbital ephemeris reported in Table~\ref{tab:solution} and taking into account the spin frequency evolution during the outburst. The best-fitting model (red line) is the superposition of four sinusoidal functions with harmonically related periods. For clarity, we show two cycles of the pulse profile. }
\label{fig:prof}
\end{figure}

\begin{figure}
\centering
\includegraphics[width=0.48\textwidth]{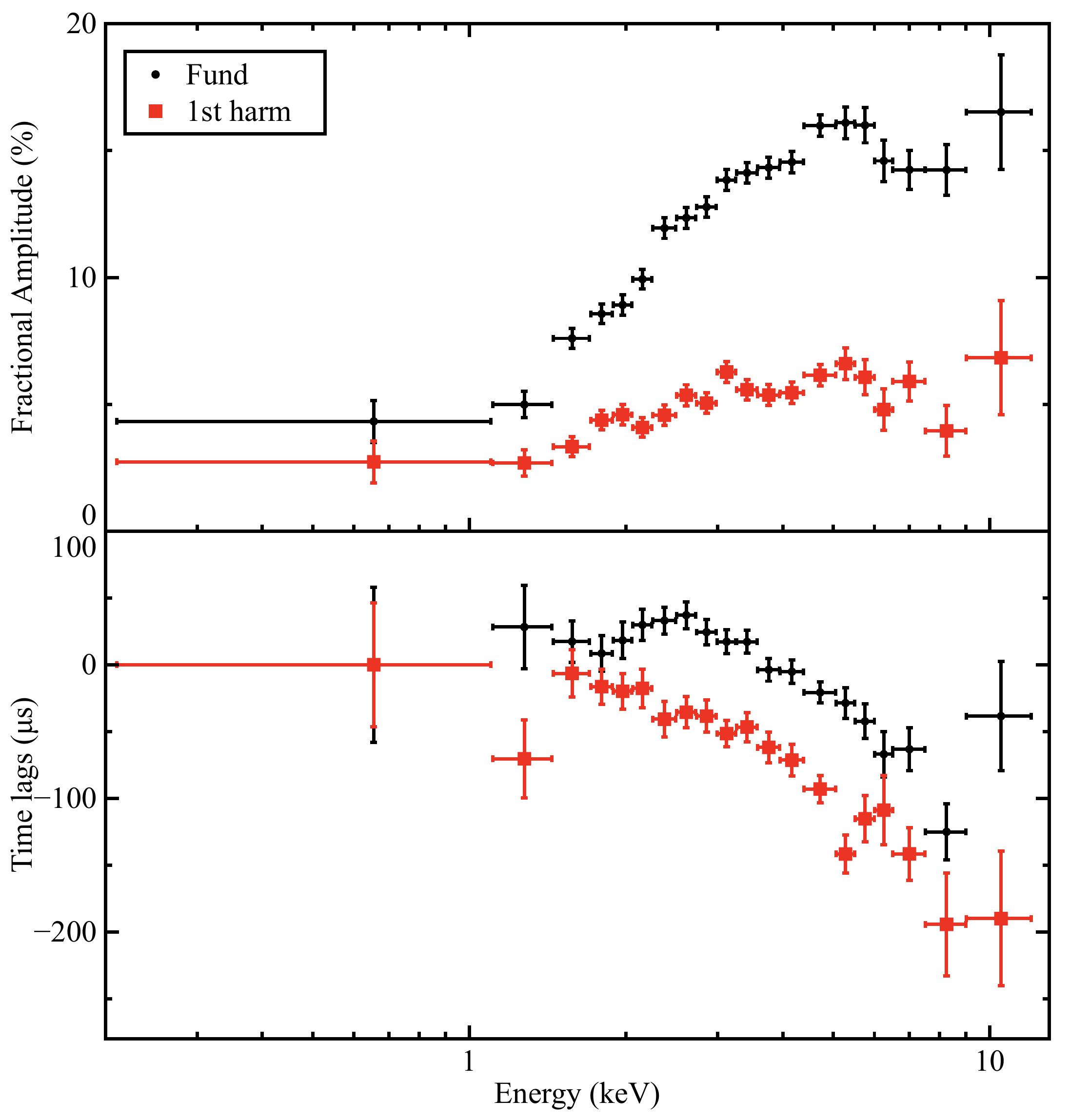}
\caption{{\it Top panel:} Pulse profile fractional amplitude evolution as a function of energy of the fundamental (black points) and first harmonic (red points) components used to model the pulse profile obtained combining the whole \nicer{} dataset. {\it Bottom panel:} Time lags in $\mu$s as a function of energy, calculated for the fundamental (black points) and first harmonic (red points) components with respect to the first energy band (0.2-1.1 keV).}
\label{fig:amp_vs_energy}
\end{figure}

\begin{figure*}
\centering
\includegraphics[width=0.8\textwidth]{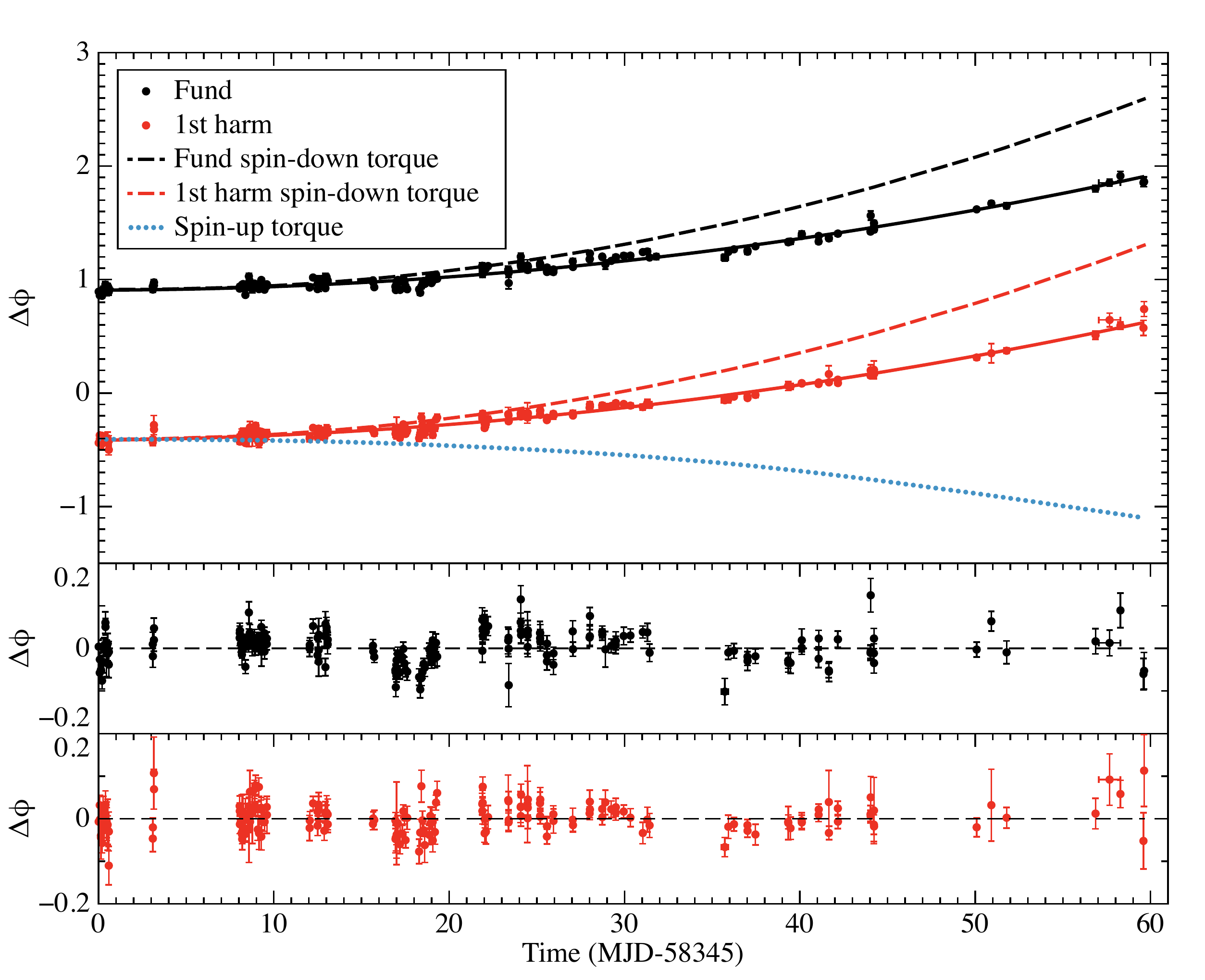}
\caption{\textit{Top panel -} Time evolution of the phase delays for the spin frequency used created by epoch-folding segments of \nicer{} data with duration between 200 and 1500 seconds. The solid lines represent the best-fitting model, while the dashed red line and the dotted cyan line represent the spin-down and the spin-up torque components used to create the magnetic disc threaded model used to fit the data. The dot-dot-dashed blue line represents the asymptotic linear trend of the spin-up torque model expected at the end of the outburst when the mass accretion rate drops. \textit{Middle and bottom panel -} Residuals in pulse phase units with respect to its best-fitting model.}
\label{fig:torque}
\end{figure*}

\section{Discussion and Conclusions}

Taking advantage of the extensive monitoring of the AMXP \igr{} performed by \nicer{}, we investigated the NS spin frequency evolution by performing coherent timing analysis during the whole outburst of the source, that lasted slightly longer than 2 months. We obtained an updated set of orbital parameters which are compatible within the errors with the previous timing solution obtained from the analysis of the first 10 days of the outburst \citep{Sanna:2018ab}.

\subsection{Overall pulse profile}

After correcting the photon time of arrivals for the timing solution reported in Tab.~\ref{tab:solution}, we epoch-folded the whole \nicer{} dataset of the source into a 32 bins pulse profile. As shown in Fig.~\ref{fig:prof}, the best-fitting model for the profile requires a combination of four sinusoids, where the fundamental, second, third, and fourth harmonics have fractional amplitudes of $~11\%$, $~4.8\%$, $~0.7\%$, $~0.16\%$, respectively. Harmonically rich pulse profiles have been reported in several other AMXPs, such as SWIFT J1749.4$-$2807 \citep{Altamirano2011a}, IGR J17511$-$3057 \citep{Riggio11a}, IGR J17379$-$3747 \citep{Sanna:2018ac} and IGR J16597$-$3704 \citep{Sanna2018aa}.

\subsection{Spin frequency evolution}

As reported in the previous section, during the outburst the pulse phase clearly shows significant evolution relative to a constant spin frequency model. More specifically, when modelled with a second order polynomial function, the pulse phase delays revealed a significant spin frequency derivative describing a deceleration of the NS (spin-down) of the order of $\sim -7\times 10^{-14}$ Hz/s. This result is confirmed independently from the timing analysis of both the fundamental and first harmonic components, for which the values of spin-down derivative are compatible, within errors, with each other. The detection of the second (or higher) harmonic component during the whole outburst has been reported also for other AMXPs such as SWIFT J1749.4$-$2807 \citep{Altamirano2011a}, XTE J1814-338 \citep{Papitto07}, IGR J17511$-$3057 \citep{Riggio11a}, SAX J1808.4$-$3658 \cite[see e.g.][]{Burderi06,Hartman08}, SWIFT J1756.9-2508 \citep{Bult2018aa} and XTE J1807$-$294 \cite[see e.g.][]{Riggio08,Patruno2010a}.      

Four other AMXPs show spin-down during their outbursts: XTE J0929$-$314 with a spin frequency derivative of $\dot{\nu}=-9.2(4)\times 10^{-14}$ Hz s$^{-1}$ \citep{Galloway02}; XTE J1814$-$338 with a spin frequency derivative of $\dot{\nu}=-6.7(7)\times 10^{-14}$ Hz s$^{-1}$ \citep{Papitto07}; IGR J17498$-$2921 with a spin frequency derivative of $\dot{\nu}=-6.3(1.9)\times 10^{-14}$ Hz s$^{-1}$ \citep{Papitto11c} and SAX J1808.4$-$3658 for which evidences of a spin-down derivative have been reported during the decay phase of its 2002 outburst \citep[$\dot{\nu}=-7.6(1.5)\times10^{-14}$ Hz s$^{-1}$;][]{Burderi06} and during its 2019 outburst \citep[$\dot{\nu}=-3.0(1)\times 10^{-13}$ Hz s$^{-1}$;][however, see the discussion for possible interpretations of the result]{Bult19}. It is noteworthy that the observed values of spin-down derivative are similar, despite the different outburst properties in which they have been detected as well as the wide range of NS spin frequencies of the AMXPs showing this phenomenon.

\subsection{Torques acting onto the NS}

From a very general point of view, the torques acting onto a NS with a significant dipolar magnetic field could be computed by adopting few general prescriptions to describe the interaction of the accreting matter with the NS. In particular: matter is accreting through a Keplerian accretion disc, truncated at the radius $R_{\rm m}$ at which matter is forced, by the action of the magnetic field, to hook onto the magnetic field lines, sharing their rigid rotation motion at the NS angular frequency. This condition implies that accretion is centrifugally inhibited if $R_{\rm m} \gg R_{\rm CO} = 
(GM/\Omega^2)^{1/3}$ where $G$ and $M$ are the gravitational constant and the NS mass, respectively, and $\Omega = 2 \pi \nu$ is the angular frequency of the NS. 

Matter accreting onto the NS surface generate a material torque determined by the accretion of its angular momentum onto the NS, $\tau_{\rm + \, \dot{M}}$, where $\dot{M}$ is the mass accretion rate. In particular, the NS is surrounded by a Keplerian accretion disc, whose innermost rim is defined by $R_{\rm m}$ and it is accreting the specific Keplerian angular momentum $\ell = \sqrt{GMR_{\rm m}}$ at $R_{\rm m}$ through a material torque:
\begin{equation}
\label{eq:taumat} 
\tau_{+ \, \dot{M}} = \ell \dot{M} = \sqrt{GMR_{\rm m}} \times \dot{M}.
\end{equation}

In the most general case of a Keplerian disc that is threaded by the magnetic field lines \cite[see e.g.][]{Ghosh79a,Wang87,Wang95,Wang96, Rappaport04, Kluzniak07}, the interaction of the orbiting matter with the magnetic filed lines (rigidly connected with the NS) determine the rise of two torques of opposite signs. A positive torque $\tau_{\rm + \, B}$ generates (we choose to use the plus sign to represent torques acting concordantly with the material torque) from interaction of the orbiting matter with the magnetic filed lines at radii $R_{\rm m} \le r \le R_{\rm CO}$, while a negative torque 
$\tau_{\rm - \, B}$, arises from the interaction of the orbiting matter with the magnetic filed lines at radii $R_{\rm CO} \le r \le R_{\rm LC}$, where $R_{\rm LC} = c/\Omega$ is the light-cylinder radius, beyond which the magneto-static structure of the field is truncated and a radiative solution is present (with $c$ being the speed of light). 

To compute the NS torque associated with this radiative solution, we adopted, as a reasonable first order guess, that Larmor's formula holds for the NS regarded as a magneto-dipole rotator. In this case, the radiative solution beyond $R_{\rm LC}$ implies a negative torque acting onto the NS: 
\begin{equation}
\label{eq:taur1} 
\tau_{\rm - \, RAD} \sim - 2/(3 c^3) (\mu \sin i )^2 \Omega^3,
\end{equation}
where $\mu = B_{\rm eq} R^3$ is the magnetic moment of the NS computed from the value of the surface magnetic field at the NS equator, $B_{\rm eq}$, $R$ is the NS radius and $i$ is the angle between the magnetic dipole moment and the spin axis of the NS. 

To compute the torque associated to the interaction of the orbiting matter with the magnetic filed lines, it is reasonable to assume that $\tau_{\rm \pm \, B}$ results from the integration of a \textit{torque radial density},
$d\tau_{\rm \pm \, B}/dr$,  from $R_{\rm CO}$ to $R_{\rm LC}$. 
The \textit{torque radial density} is determined by specific prescriptions for the structure of the magnetic field in the mid-plane of the accretion disc. In the following, in line with \citet{Kluzniak2007}, we adopted a quite simple prescription
for the ratio of the toroidal to poloidal components of the magnetic field threading the accretion disc at a generic radius $r$:
\begin{equation} 
\label{eq:bfield1}
\frac{B_{\phi}(r)}{B_{z}(r)} = 1 - \frac{\Omega_{\rm K}(r)}{\Omega}.
\end{equation}
In line with current wisdom, the poloidal magnetic field is a \textit{screened} version of the NS dipolar field at the disc mid-plane, $B_{z}(r) = \eta  \mu/r^3$, with $\eta \sim 1$
\footnote{On the other hand, several authors \citep[see e.g.][and references therein]{Wang87,Rappaport04,Kluzniak2007} discussed different prescription for the magnetic field structure like the following \citep[computed in detail in][]{Kluzniak2007}:
\begin{equation} 
\label{eq:bfield2}
\frac{B_{\phi}(r)}{B_{z}(r)} =
\left\{
\begin{array}{ll}
1 - \Omega_{\rm K}(r)/\Omega & {\; \; \; \rm for \,} r > R_{\rm CO} \\
 & \\
\Omega/\Omega_{\rm K}(r) - 1 & {\; \; \; \rm for \,} r \le R_{\rm CO} \\
\end{array}
\right.
\end{equation}
and other more sophisticated and physically motivated versions of the relations above \citep[see the exhaustive discussion in][]{Kluzniak2007}. However, within factors of the order unity, the final results on the torques are quite independent of the detailed prescription adopted for the magnetic field, particularly for $R_{\rm m} \rightarrow R_{\rm CO}$ 
\citep{Kluzniak2007} which will be the case for \igr{}, as we verified {\it a posteriori} (see below).}.

Let us consider $\tau_{\rm - \, B}$, resulting from the integration of the torque radial density from $R_{\rm CO}$ to $R_{\rm LC}$. Since the dipolar field decreases as $r^{-3}$, it is reasonable to expect \cite[see e.g.][]{Rappaport04} that the integral is dominated by the contribution close to the corotation radius and therefore independent of 
$R_{\rm LC}$. According to \citet[][]{Rappaport04} the computation of this term gives the expression
\begin{equation}
\label{eq:taubm} 
\tau_{\rm - \, B} = -  \frac{\mu^2}{9R_{\rm CO}^3}.
\end{equation} 
Finally, after some algebraic manipulation, $\tau_{\rm - \, RAD}$ can be expressed in terms of $\tau_{\rm - \, B}$ as
\begin{equation}
\tau_{\rm - \, RAD} = 6 (\sin i)^2\left(\frac{R_{\rm CO}}{R_{\rm LC}}\right)^3 \tau_{\rm - \, B}.
\end{equation}
Since $R_{\rm CO}$ and $R_{\rm LC}$ are fixed for a given NS spin frequency, the term $\tau_{\rm - \, RAD}$ can be absorbed in the $\tau_{\rm - \, B}$ term as:
\begin{equation}
\label{eq:taum} 
\tau_{\rm - \, B} + \tau_{\rm - \, RAD} = - \gamma\frac{\mu^2}{9R_{\rm CO}^3},
\end{equation} 
where $\gamma = 1 + 6 (\sin i)^2(R_{\rm CO}/R_{\rm LC})^3 \sim 1$, since, typically, $R_{\rm CO}/R_{\rm LC} \ll 1$.

Using an explicit prescription for the torque radial density, as derived from a specific prescription 
for the magnetic field structure, the positive torque $\tau_{\rm + \, \dot{M}} + \tau_{\rm + \, B}$ can be computed as
\begin{equation}
\label{eq:taup} 
\tau_{\rm + \, B} =  \dot{M} + \int_{R_{\rm m}}^{R_{\rm CO}} \left( \frac{d\tau_{\rm + \, B}}{dr}\right) dr.
\end{equation}

In summary, we adopted the general prescription for the total torque exerted on the NS as derived in \citet[][Eq.~ 36 of the paper]{Kluzniak2007}, modified by the introduction of the factor $\gamma = 1 + 6 (\sin i)^2(R_{\rm CO}/R_{\rm LC})^3 \sim 1$ to take into account the torque from the radiative solution. 
Defining $\tau_{\rm TOT} = \tau_{+ \, \dot{M}} + \tau_{+ \, B} + \tau_{- \, B} + \tau_{\rm - \, RAD} $, we find:
\begin{equation} 
\label{eq:tautot}
\tau_{\rm TOT} = \dot{M} \sqrt{GMR_{\rm m}} - \gamma \frac{\mu^2}{9R_{\rm m}^3}\left( 3 - 2 \sqrt{\frac{R_{\rm CO}^3}{R_{\rm m}^3}}\right),
\end{equation}
which depends on the exact location of the truncation radius $R_{\rm m}$
\footnote{Adopting different prescriptions for the torque radial density, as derived from a specific model 
for the magnetic field structure, the functional prescription for the total torque varies slightly. In particular, \citep{Rappaport04} and \citep{Kluzniak2007} computed the total torque acting on the NS for the prescription defined in equation (\ref{eq:bfield2}).
The final torque is reported in Eq.~B13 \citep[APPENDIX B;][]{Kluzniak2007}  and in Eq.~24 \citep{Rappaport04}. We note that small typos are probably present in the latter equation, although the limit for $R_{\rm m} \rightarrow R_{\rm CO}$, namely $\tau_{\rm TOT} = \dot{M} \sqrt{GMR_{\rm CO}} - \gamma \mu^2/(9R_{\rm CO}^3)$ is the same for the two expressions,
as well as for Eq.~\ref{eq:tautot} adopted in this work. Indeed since we will find {\it a posteriori} that the condition
$R_{\rm m} \rightarrow R_{\rm CO}$ was applicable during the entire 2018 outburst of \igr{}, we are confident that our results are quite independent of the detailed prescription adopted for the magnetic field structure.}. 
Detailed computations of the truncation radius has been performed by several authors by carefully balancing the torques acting in the disc and requiring that, at the truncation radius, viscous torques vanishes \citep[see e.g.][and references therein]{Wang95,Kluzniak2007}. 
Defining the Alfv\'en radius as:
\begin{equation}
\label{eq:alfven} 
R_{\rm A} = (GM)^{-1/7} \mu^{4/7} \dot{M}^{-2/7},
\end{equation}
the location of $R_{\rm m}$ depends on the ratio $R_{\rm A}/R_{\rm CO} = \xi$ \citep[see e.g.][]{Wang95,Kluzniak2007} being:
\begin{equation} 
\label{eq:trunc}
R_{\rm m} = R_{\rm CO} \times 
\left\{
\begin{array}{ll}
2^{1/5} \xi^{7/10} & {\; \; \; \rm for \,}  R_{\rm A} \ll R_{\rm CO} \\
 & \\
1 - \xi^{-7/2} & {\; \; \; \rm for \,} R_{\rm A} \gg R_{\rm CO} \\
\end{array}
\right.
\end{equation}
For a reasonable range of Alfv\'en radii, namely $0.25 R_{\rm CO} \leq R_{\rm A} \leq 10 R_{\rm CO}$, the results 
of Eq.~\ref{eq:trunc} can be approximated \citep[from a visual inspection of Fig.~1 of][]{Kluzniak2007} with
\begin{equation} 
\label{eq:trap}
R_{\rm m} \sim 
\left\{
\begin{array}{ll}
R_{\rm A} & {\; \; \; \rm for \,}  R_{\rm A} < R_{\rm CO} \\
 & \\
R_{\rm CO} & {\; \; \; \rm for \,} R_{\rm A} \gtrsim R_{\rm CO}. \\
\end{array}
\right.
\end{equation}
In general, in the range $0.25 R_{\rm CO} \leq R_{\rm A} \leq 10 R_{\rm CO}$, Eq.~\ref{eq:trap} can be 
assumed as a reasonable estimate of the truncation radius for several prescriptions of the magnetic field structure.
Inserting Eq.~\ref{eq:trap} in Eq.~\ref{eq:tautot}, we find the correct prescription for the torque acting on the NS.
In particular, if, throughout the outburst of \igr{}, the condition 
$R_{\rm A} \gtrsim R_{\rm CO}$ was met, the torque is:
\begin{equation} 
\label{eq:taut2}
\tau_{\rm TOT} = \dot{M} \sqrt{GMR_{\rm CO}} 
- \gamma \frac{\mu^2}{9R_{\rm CO}^3}
\end{equation}
Conservation of angular momentum therefore implies:
\begin{equation}
\label{eq:torque} 
\tau_{\rm TOT} = 2 \pi I_{\rm NS} \dot{\nu} = \dot{M}(t) \sqrt{GMR_{\rm CO}} - \gamma \frac{\mu^2}{9R_{\rm CO}^3}, 
\end{equation} 
where $I_{\rm NS}$ is the NS moment if inertia (for simplicity, here, we assumed that the moment of inertia of the NS does not vary in response to the accretion process, although this small effect can be taken into account if a reliable Equation of State for the NS is assumed\footnote{More in detail let us assume that the NS moment of inertia has the form $I_{\rm NS} = \zeta(n) (2/5)MR^2$, where
$\zeta(n) \sim 1$ is a constant that depends, mainly, on the polytropic index $n$ adopted to describe the NS. Logarithmic differentiation and some algebraic manipulation gives for the NS angular momentum: $dL_{\rm NS}/dt = 2 \pi I_{\rm NS} \dot{\nu} + \dot{M} \Omega R^2 (2\zeta(n)/5)[1+2(\dot{R}/\dot{M})(M/R)]$. Adopting $2\zeta(n)/5 \sim 1$ and 
$(\dot{R}/\dot{M})(M/R) \sim 0$ (since in most of the Equation of State of the ultra-dense matter the radius is almost independent of the NS mass) we find $dL_{\rm NS}/dt \simeq 2 \pi I_{\rm NS}\dot{\nu} + \dot{M} \Omega R^2$. This 
last term has been included in \citet[][Eq.~36 and Eq.~B13 of that paper]{Kluzniak2007}.
It is worth noting that Eq.~\ref{eq:torque} is correct only for negligible variation of $I_{\rm NS}$ during the accretion process, condition that can be easily verified for this work using Eq.~7 in \citet{Sanna:2017aa}}, and $\dot{\nu}$ is the NS spin frequency derivative.

If the mass accretion rate is constant, the net torque in Eq.~\ref{eq:torque} is constant as well, positive or negative depending on the relative strengths of the acting torques. 
On the other hand, when $\dot{M}(t)$ varies, which is the case during an outburst, variations of the accretion rate must be taken into account when computing the spin frequency derivative. 
Under the assumption the mass accretion rate is well described by the bolometric accretion luminosity $L_{\rm BOL}$, we can relate the latter quantity to the observed source flux within the \nicer{} energy range. We can then define $L_{\rm BOL}= 4\pi\,d^2\kappa_{\rm BOL} \Phi_{\rm 0.5-10 \, keV}$, where $d$ is the source distance and $\kappa_{\rm BOL} \geq 1$ is a constant, depending on the spectral shape. Finally, the bolometric accretion luminosity is related to the mass accretion rate through the efficiency $\eta_{\rm ACC}$, and therefore:
\begin{equation}
\label{eq:mdot} 
\dot{M}(t) = \left( \frac{\kappa_{\rm BOL} 4 \pi R}{GM\eta_{\rm ACC}} \right)  d^2\Phi_{\rm 0.5-10 keV}(t) = \epsilon d^2\Phi_{\rm 0.5-10 keV}(t).
\end{equation} 
In summary, we can express the frequency derivative as:
\begin{equation}
\label{eq:nudot} 
\dot{\nu}(t) =\frac{1}{2 \pi I_{\rm NS}}\Big[ \Omega R_{\rm CO}^2 \, \epsilon d^2 \Phi_{\rm 0.5-10 keV}(t) - \gamma \frac{\mu^2}{9R_{\rm CO}^3}\Big]
\end{equation} 
where we used the identity $\sqrt{GMR_{\rm CO}} = \Omega R_{\rm CO}^2$.

In recent years, variations of the spin frequency derivative caused by the temporal evolution of the source flux $\phi(t)$ during an outburst were modelled assuming simple analytic prescriptions for the flux evolution \cite[see e.g. IGR J00291+5934 and IGR J17511-3057][]{Burderi07,Riggio11a}. More recently, a numerical approach in which the torque is rescaled at every time using the instantaneous value of the X-ray flux has been developed and applied successfully to the timing of the X-ray pulsar GRO J1744-28, a slowly spinning ($2.14 \, {\rm Hz}$) NS accreting from a low-mass companion \citep{Sanna:2018ab}.  Thanks to the excellent \nicer{} monitoring campaign that furnished a reliable and well sampled (1 point every one or two days) $\rm 0.5-10 \, keV$ flux curve of the \igr{}, we were able to apply, for the first time, this technique to an AMXP. 

Starting from Eq.~\ref{eq:ph}, we replaced the factor $\frac{1}{2}\dot{\nu}\,(t-T_0)^2$ with the torque-induced phase delay term:

\begin{eqnarray}
\Delta \phi_{\tau}(t)&=& \Delta \phi_{\tau M}(t)+\Delta \phi_{\tau B}(t) =\\
			&=&\frac{1}{2\pi I}\int_{T_0}^tdt'\int_{T_0}^{t'} \Omega R_{\rm CO}^2 \, \epsilon d^2 \Phi_{\rm 0.5-10 keV}(t)dt''+\\
			&-&\frac{1}{2\pi I}\int_{T_0}^tdt'\int_{T_0}^{t'} \gamma \frac{\mu^2}{9R_{\rm CO}^3}dt'',
\end{eqnarray}
where $\Delta \phi_{\tau M}(t)$ and $\Delta \phi_{\tau B}(t)$ represent the contributions from the material and the threading torques, respectively. To be able to integrate the source flux in the aforementioned relation, we applied cubic spline data interpolation techniques to generate the outburst flux profile starting from that reported in Fig.~\ref{fig:flux}.
Using this new prescription, we then modelled simultaneously the pulse phase delays from the fundamental and first harmonic components created following the procedure described in the previous section. More specifically, we linked the model components $R_{orb}(t)$ and $\Delta \phi_{\tau M}(t)$ between the two sets of data. We iterated the method, until no significant differential corrections were found for the parameters of the model. To reduce the possible correlation between the model parameters, we limited the exploration range for the material torque parameter by defining a meaningful interval for the constants included in Eq.~\ref{eq:mdot}. More specifically, we limited the accretion efficiency in the range $0.75<\eta_{\rm ACC}<1$ in agreement with some recently proposed estimates (Burderi et al., 2020, submitted), we considered the source located at $d=7.6\pm0.7$ kpc \cite[obtained from the observation of a photo-spheric radius expansion type-I X-ray burst observed by INTEGRAL during the outburst][]{Kuiper2020} and we considered the bolometric conversion factor with respect to the unabsorbed flux in the 0.5-10 keV energy range bounded between $2<\kappa_{\rm BOL}<3$. To determine the latter interval, we used the broad-band (0.5-80 keV) energy band spectra obtained combining the simultaneous Swift/XRT, NuSTAR and INTEGRAL IBIS/ISGRI observations of \igr{} at the beginning of the outburst \citep{Sanna:2018ab} to determine the ratio between the low (0.5-10 keV) and high energy (0.5-80 keV) bands obtaining a factor $\sim 3$. Moreover, we estimated the same constant, at different stages of the outburst, by using the 0.5-80 keV unabsorbed flux extrapolated from the best-fitting model of the available \nicer{} energy spectra from which we obtained values ranging between 2 and 3. It is worth to noting that this discrepancy is likely reflecting the limited energy coverage of the \nicer{} data, therefore, we decided to consider the largest interval (2-3) for the parameter.
 
The best-fit model parameters are reported in Tab.~\ref{tab:torque}, while the top panel of Fig.~\ref{fig:torque} shows the fundamental (black points) and first harmonic (red points) pulse phase delays, the best-fitting model (solid lines), the spin-down (dashed lines) and spin-up (dotted line) torque components. The medium and bottom panels show the delay residuals (in phase units) with respect to the best-fitting model for the frequency fundamental and first harmonic components, respectively. 

As expected, the orbital parameters obtained from this fit are fully consistent with those reported in Tab.~\ref{tab:solution}, since the long baseline of the dataset guarantees to clearly disentangle orbital phase modulation from and phase variation due to the NS spin frequency changes.  
From the best-fitting parameter of each frequency phase component we infer the dipolar magnetic moment $\mu_F\simeq4.1(4)\times10^{26}$ G\,cm$^3$ and $\mu_{1h}\simeq4.1(5)\times10^{26}$ G\,cm$^3$, for the fundamental and the first harmonic, respectively. Adopting the Friedman-Pandharipande-Skyrme (FPS) equation of
state \cite[see e.g.][] {Friedman1981a,Pandharipande1989a} for a 1.4 M$_\odot$ NS, we estimate a NS radius of $R = 1.14\times 10^6$ cm, and a moment of inertia $I\simeq1.45\times10^{45}$ g cm$^{2}$. This corresponds to an equatorial magnetic field of $B_{eq}=\mu^2/R^3=2.8(3)\times10^8\simeq 2.8(3)\times10^8$ G. This value is consistent both with the estimation reported in \citet{Sanna:2018ab} from spin frequency equilibrium considerations for the X-ray pulsar, and the average magnetic field of known AMXPs \cite[][]{Mukherjee2015}.  It is noteworthy that the value of $B_{eq}$ obtained here is likely overestimated due to a non accurate estimation of mass accretion rate in case of spin-down regime. In fact, as discussed by \citealt{Kluzniak2007}, in this regime the angular momentum deposited in the disc by the pulsar torques leads to a substantial additional dissipation of energy in the disc, hence, the mass accretion rate estimated from Eq.~\ref{eq:mdot} results on an upper limit of the quantity. However, since it is not quite clear what fraction of this additional dissipation of energy is released in non-thermal process and what is kept inside the disc, it is complicated to quantify the discrepancy. We started approaching the problem considering the worst case scenario at which the whole rotational energy of the pulsar $I\omega\dot{\omega}$ is transferred and released by the accretion disc. With this hypothesis, we subtract the rotation luminosity from observed luminosity in Eq.~\ref{eq:mdot} and we then fitted the frequency phase delays. We obtained that, taking into account the correction on the mass accretion rate, the NS magnetic field decreases roughly by 15\%, reaching the value $B_{eq}=2.4(4)\times10^8$ G. Both the properties of the source during the outburst and the magnetic field estimation obtained from the studies of the spin-down like behaviour of the phase delays do not seem to suggest \igr{} as an exception among the AMXPs.

\begin{table}

\begin{tabular}{l | c }
\hline
Parameters             & \\
\hline
\hline
$P_{orb}$ (s) &31684.7506(5)\\
a sin\textit{i/c} (lt-s) &1.227728(6)\\
$T_{\text{NOD}}$ (MJD) & 58345.1719787(10)\\
Eccentricity (e) &< 5$\times 10^{-5}$\\
$\chi^2$/d.o.f. &872.5/355 \\
\hline
\hline
Fundamental\\
\hline
\hline
Spin frequency $\nu_0$ (Hz) & 527.42570060(2)\\
$\mu$ (G\,cm$^3$) & $4.1(4)\times 10^{26}$  \\
\hline
\hline
1st Harmonic\\
\hline
\hline
Spin frequency $\nu_0$ (Hz) & 527.42570058(2)\\
$\mu$ (G\,cm$^3$) & $4.1(5)\times 10^{26}$\\
\hline
\end{tabular}
\caption{Orbital ephemerides of \igr{} obtained from the torque-fitting timing analysis of the \nicer{} observations collected during the whole outburst of the source. The orbital solution is referred to the epoch T$_0$=58345.0 MJD. Errors are at 1$\sigma$ confidence level.}
\label{tab:torque}
\end{table}

\subsection{Pulse profile energy dependence}

We investigated the properties of the pulse profile of \igr{} as a function of energy (0.2-12 keV) by considering the whole available \nicer{} dataset from its 2018 outburst. Although the overall pulse profile is well described by the superposition of four sinusoidal functions harmonically related (see Fig.~\ref{fig:prof}), while applying energy selections to the events the sensitivity to harmonic components larger than two drops significantly, we therefore, independently studied the fractional amplitude and the lags of the fundamental and first harmonic only. 

As shown in the top panel of Fig.~\ref{fig:amp_vs_energy}, the fundamental pulse fractional amplitude clearly increases with energy, varying from 4\% to 16\% up to 5 keV, followed by a decreasing trend down to $\sim 12$\% at 12 keV. The first harmonic fractional amplitude resembles the fundamental component, with a smaller variation between 2.6\% and 6.6\% up to 5 keV, though. Similar behaviour for the fractional amplitude has been already reported for other AMXPs such as SWIFT J1756.9-2508 \citep{Sanna:2018aa}, SAX J1808.4-3658 \citep{Patruno09a,Sanna2016a}, Aql X$-$1 \citep{Casella08}, XTE J1807$-$294 \citep{Kirsch04} and IGR J00291+5934 \citep{Falanga05b,Sanna2017b}. Mechanisms such as a strong Comptonisation of the beamed radiation have been applied, quite successfully, to describe the properties of a few sources \citep{Falanga07b}. However, it is noteworthy that quite different energy distributions of fractional pulse amplitude have been observed in other AMXPs such as XTE J1751$-$305 \citep{Falanga07b}, \saxj{} \citep{Cui98b,Falanga07b,Sanna:2017ab} and IGR J17511$-$3057 \citep{Papitto10,Falanga11,Riggio11a}.

Finally, time lags (Fig.~\ref{fig:amp_vs_energy}, bottom panel) of the fundamental component suggest almost no lags between pulsed photons with energy in the range 0.5-3.5 keV, while hard lags are clearly observed for higher energies. On the other hand, the first harmonic component clearly shows hard lags with increasing energies. Although within a limited energy range, these results are consistent with those reported by \citet{Falanga07b,Falanga11} for the AMXPs XTE J1751$-$305,  XTE J17511$-$3057 and SAX J1808.4$-$3658, where the hard lags have been interpreted as the result of soft X-ray photons coming from the disc or the NS surface upscatter off hot electrons in the accretion column.

\section*{Acknowledgments}

The NICER mission and portions of the NICER science team activities are funded by NASA. C.M. is supported by an appointment to the NASA Postdoctoral Program at the Marshall Space Flight Center, administered by Universities Space Research Association under contract with NASA.

\bibliographystyle{mn2e}
\bibliography{biblio}

\label{lastpage}

\end{document}